
%
\font\twelverm=cmr10 scaled 1200    \font\twelvei=cmmi10 scaled 1200
\font\twelvesy=cmsy10 scaled 1200   \font\twelveex=cmex10 scaled 1200
\font\twelvebf=cmbx10 scaled 1200   \font\twelvesl=cmsl10 scaled 1200
\font\twelvett=cmtt10 scaled 1200   \font\twelveit=cmti10 scaled 1200
\skewchar\twelvei='177   \skewchar\twelvesy='60
\def\twelvepoint{\normalbaselineskip=12.4pt
  \abovedisplayskip 12.4pt plus 3pt minus 9pt
  \belowdisplayskip 12.4pt plus 3pt minus 9pt
  \abovedisplayshortskip 0pt plus 3pt
  \belowdisplayshortskip 7.2pt plus 3pt minus 4pt
  \smallskipamount=3.6pt plus1.2pt minus1.2pt
  \medskipamount=7.2pt plus2.4pt minus2.4pt
  \bigskipamount=14.4pt plus4.8pt minus4.8pt
  \def\rm{\fam0\twelverm}          \def\it{\fam\itfam\twelveit}%
  \def\sl{\fam\slfam\twelvesl}     \def\bf{\fam\bffam\twelvebf}%
  \def\mit{\fam 1}                 \def\cal{\fam 2}%
  \def\tt{\twelvett}
  \textfont0=\twelverm   \scriptfont0=\tenrm   \scriptscriptfont0=\sevenrm
  \textfont1=\twelvei    \scriptfont1=\teni    \scriptscriptfont1=\seveni
  \textfont2=\twelvesy   \scriptfont2=\tensy   \scriptscriptfont2=\sevensy
  \textfont3=\twelveex   \scriptfont3=\twelveex  \scriptscriptfont3=\twelveex
  \textfont\itfam=\twelveit
  \textfont\slfam=\twelvesl
  \textfont\bffam=\twelvebf \scriptfont\bffam=\tenbf
  \scriptscriptfont\bffam=\sevenbf
  \normalbaselines\rm}

\def\beginlinemode{\endmode
  \begingroup\parskip=0pt \obeylines\def\\{\par}\def\endmode{\par\endgroup}}
\def\beginparmode{\endmode
  \begingroup \def\endmode{\par\endgroup}}
\let\endmode=\par
{\obeylines\gdef\
{}}
\def\singlespace{\baselineskip=\normalbaselineskip}

\def\oneandahalfspace{\baselineskip=\normalbaselineskip
  \multiply\baselineskip by 3 \divide\baselineskip by 2}
\def\doublespace{\baselineskip=\normalbaselineskip \multiply\baselineskip by 2}
\newcount\firstpageno
\firstpageno=2
\footline={\ifnum\pageno<\firstpageno{\hfil}%
\else{\hfil\twelverm\folio\hfil}\fi}
\let\rawfootnote=\footnote              
\def\footnote#1#2{{\rm\singlespace\parindent=0pt\rawfootnote{#1}{#2}}}
\def\raggedcenter{\leftskip=2em plus 12em \rightskip=\leftskip
  \parindent=0pt \parfillskip=0pt \spaceskip=.3333em \xspaceskip=.5em
  \pretolerance=9999 \tolerance=9999
  \hyphenpenalty=9999 \exhyphenpenalty=9999 }
\parskip=\medskipamount
\twelvepoint            
\overfullrule=0pt       
\def\preprintno#1{
 \rightline{\rm #1}}    
\def\author                     
  {\vskip 3pt plus 0.2fill \beginlinemode
   \singlespace \raggedcenter \twelvesc}
\def\affil                      
  {\vskip 3pt plus 0.1fill \beginlinemode
   \oneandahalfspace \raggedcenter \sl}
\def\abstract                   
  {\vskip 3pt plus 0.3fill \beginparmode
   \doublespace \narrower \noindent ABSTRACT: }
\def\endtitlepage               
  {\endpage                     
   \body}
\def\body                       
  {\beginparmode}               

\def\subhead#1{                 
  \vskip 0.12truein             
  {\raggedcenter #1 \par}
   \nobreak\vskip 0.1truein\nobreak}
\def\refto#1{$|{#1}$}           
\def\references                
  {\subhead{References}        
   \beginparmode
   \frenchspacing \parindent=0pt \leftskip=1truecm
   \parskip=8pt plus 3pt \everypar{\hangindent=\parindent}}
\gdef\refis#1{\indent\hbox to 0pt{\hss#1.~}}    
\gdef\journal#1, #2, #3, 1#4#5#6{               
    {\sl #1~}{\bf #2}, #3, (1#4#5#6)}           
\def\refstylenp{                
  \gdef\refto##1{ [##1]}                                
  \gdef\refis##1{\indent\hbox to 0pt{\hss##1)~}}        
  \gdef\journal##1, ##2, ##3, ##4 {                     
     {\sl ##1~}{\bf ##2~}(##3) ##4 }}
\def\refstyleprnp{              
  \gdef\refto##1{ [##1]}                                
  \gdef\refis##1{\indent\hbox to 0pt{\hss##1)~}}        
  \gdef\journal##1, ##2, ##3, 1##4##5##6{               
    {\sl ##1~}{\bf ##2~}(1##4##5##6) ##3}}
\def\pr{\journal Phys. Rev., }

\def\prl{\journal Phys. Rev. Lett., }
\def\prpts{\journal Phys. Rep., }
\def\np{\journal Nucl. Phys., }
\def\pl{\journal Phys. Lett., }

\def\endreferences{\body}
\def\endpage                    
  {\vfill\eject}
\def\endpaper                   
  {\endmode\vfill\supereject}
\def\endit
  {\endpaper\end}
\def\ref#1{Ref. #1}                     
\def\Ref#1{Ref. #1}                     

\def\m@th{\mathsurround=0pt }
\font\twelvesc=cmcsc10 scaled 1200
\def\cite#1{{#1}}
\def\(#1){(\call{#1})}
\def\call#1{{#1}}
\def\taghead#1{}
\def\leaderfill{\leaders\hbox to 1em{\hss.\hss}\hfill}
\def\twiddle{\lower.9ex\rlap{$\kern-.1em\scriptstyle\sim$}}
\def\bigtwiddle{\lower1.ex\rlap{$\sim$}}
\def\gtwid{\mathrel{\raise.3ex\hbox{$>$\kern-.75em\lower1ex\hbox{$\sim$}}}}
\def\ltwid{\mathrel{\raise.3ex\hbox{$<$\kern-.75em\lower1ex\hbox{$\sim$}}}}
\def\square{\kern1pt\vbox{\hrule height 1.2pt\hbox{\vrule width 1.2pt\hskip 3pt
   \vbox{\vskip 6pt}\hskip 3pt\vrule width 0.6pt}\hrule height 0.6pt}\kern1pt}
\catcode`@=11
\newcount\tagnumber\tagnumber=0

\immediate\newwrite\eqnfile
\newif\if@qnfile\@qnfilefalse
\def\write@qn#1{}
\def\writenew@qn#1{}
\def\w@rnwrite#1{\write@qn{#1}\message{#1}}
\def\@rrwrite#1{\write@qn{#1}\errmessage{#1}}

\def\taghead#1{\gdef\t@ghead{#1}\global\tagnumber=0}
\def\t@ghead{}

\expandafter\def\csname @qnnum-3\endcsname
  {{\t@ghead\advance\tagnumber by -3\relax\number\tagnumber}}
\expandafter\def\csname @qnnum-2\endcsname
  {{\t@ghead\advance\tagnumber by -2\relax\number\tagnumber}}
\expandafter\def\csname @qnnum-1\endcsname
  {{\t@ghead\advance\tagnumber by -1\relax\number\tagnumber}}
\expandafter\def\csname @qnnum0\endcsname
  {\t@ghead\number\tagnumber}
\expandafter\def\csname @qnnum+1\endcsname
  {{\t@ghead\advance\tagnumber by 1\relax\number\tagnumber}}
\expandafter\def\csname @qnnum+2\endcsname
  {{\t@ghead\advance\tagnumber by 2\relax\number\tagnumber}}
\expandafter\def\csname @qnnum+3\endcsname
  {{\t@ghead\advance\tagnumber by 3\relax\number\tagnumber}}

\def\equationfile{%
  \@qnfiletrue\immediate\openout\eqnfile=\jobname.eqn%
  \def\write@qn##1{\if@qnfile\immediate\write\eqnfile{##1}\fi}
  \def\writenew@qn##1{\if@qnfile\immediate\write\eqnfile
    {\noexpand\tag{##1} = (\t@ghead\number\tagnumber)}\fi}
}

\def\callall#1{\xdef#1##1{#1{\noexpand\call{##1}}}}
\def\call#1{\each@rg\callr@nge{#1}}

\def\each@rg#1#2{{\let\thecsname=#1\expandafter\first@rg#2,\end,}}
\def\first@rg#1,{\thecsname{#1}\apply@rg}
\def\apply@rg#1,{\ifx\end#1\let\next=\relax%
\else,\thecsname{#1}\let\next=\apply@rg\fi\next}

\def\callr@nge#1{\calldor@nge#1-\end-}
\def\callr@ngeat#1\end-{#1}
\def\calldor@nge#1-#2-{\ifx\end#2\@qneatspace#1 %
  \else\calll@@p{#1}{#2}\callr@ngeat\fi}
\def\calll@@p#1#2{\ifnum#1>#2{\@rrwrite{Equation range #1-#2\space is bad.}
\errhelp{If you call a series of equations by the notation M-N, then M and
N must be integers, and N must be greater than or equal to M.}}\else%
{\count0=#1\count1=#2\advance\count1 by1\relax\expandafter\@qncall\the\count0,%
  \loop\advance\count0 by1\relax%
    \ifnum\count0<\count1,\expandafter\@qncall\the\count0,%
  \repeat}\fi}

\def\@qneatspace#1#2 {\@qncall#1#2,}
\def\@qncall#1,{\ifunc@lled{#1}{\def\next{#1}\ifx\next\empty\else
  \w@rnwrite{Equation number \noexpand\(>>#1<<) has not been defined yet.}
  >>#1<<\fi}\else\csname @qnnum#1\endcsname\fi}

\let\eqnono=\eqno
\def\eqno(#1){\tag#1}
\def\tag#1$${\eqnono(\displayt@g#1 )$$}

\def\aligntag#1\endaligntag
  $${\gdef\tag##1\\{&(##1 )\cr}\eqalignno{#1\\}$$
  \gdef\tag##1$${\eqnono(\displayt@g##1 )$$}}

\def\eqalignno#1{\displ@y \tabskip\centering
  \halign to\displaywidth{\hfil$\displaystyle{##}$\tabskip\z@skip
    &$\displaystyle{{}##}$\hfil\tabskip\centering
    &\llap{$\displayt@gpar##$}\tabskip\z@skip\crcr
    #1\crcr}}

\def\displayt@gpar(#1){(\displayt@g#1 )}

\def\displayt@g#1 {\rm\ifunc@lled{#1}\global\advance\tagnumber by1
        {\def\next{#1}\ifx\next\empty\else\expandafter
        \xdef\csname @qnnum#1\endcsname{\t@ghead\number\tagnumber}\fi}%
  \writenew@qn{#1}\t@ghead\number\tagnumber\else
        {\edef\next{\t@ghead\number\tagnumber}%
        \expandafter\ifx\csname @qnnum#1\endcsname\next\else
        \w@rnwrite{Equation \noexpand\tag{#1} is a duplicate number.}\fi}%
  \csname @qnnum#1\endcsname\fi}

\def\ifunc@lled#1{\expandafter\ifx\csname @qnnum#1\endcsname\relax}

\let\@qnend=\end\gdef\end{\if@qnfile
\immediate\write16{Equation numbers written on []\jobname.EQN.}\fi\@qnend}

\catcode`@=12
\catcode`@=11
\newcount\r@fcount \r@fcount=0
\def\refreset{\global\r@fcount=0}
\newcount\r@fcurr
\immediate\newwrite\reffile
\newif\ifr@ffile\r@ffilefalse
\def\w@rnwrite#1{\ifr@ffile\immediate\write\reffile{#1}\fi\message{#1}}

\def\writer@f#1>>{}
\def\referencefile{
  \r@ffiletrue\immediate\openout\reffile=\jobname.ref%
  \def\writer@f##1>>{\ifr@ffile\immediate\write\reffile%
    {\noexpand\refis{##1} = \csname r@fnum##1\endcsname = %
     \expandafter\expandafter\expandafter\strip@t\expandafter%
     \meaning\csname r@ftext\csname r@fnum##1\endcsname\endcsname}\fi}%
  \def\strip@t##1>>{}}

\def\citeall#1{\xdef#1##1{#1{\noexpand\cite{##1}}}}
\def\cite#1{\each@rg\citer@nge{#1}}

\def\each@rg#1#2{{\let\thecsname=#1\expandafter\first@rg#2,\end,}}
\def\first@rg#1,{\thecsname{#1}\apply@rg}	
\def\apply@rg#1,{\ifx\end#1\let\next=\relax
\else,\thecsname{#1}\let\next=\apply@rg\fi\next}

\def\citer@nge#1{\citedor@nge#1-\end-}	
\def\citer@ngeat#1\end-{#1}
\def\citedor@nge#1-#2-{\ifx\end#2\r@featspace#1 
  \else\citel@@p{#1}{#2}\citer@ngeat\fi}	
\def\citel@@p#1#2{\ifnum#1>#2{\errmessage{Reference range #1-#2\space is bad.}%
    \errhelp{If you cite a series of references by the notation M-N, then M and
    N must be integers, and N must be greater than or equal to M.}}\else%
{\count0=#1\count1=#2\advance\count1 by1\relax\expandafter\r@fcite\the\count0,%
  \loop\advance\count0 by1\relax
    \ifnum\count0<\count1,\expandafter\r@fcite\the\count0,%
  \repeat}\fi}

\def\r@featspace#1#2 {\r@fcite#1#2,} 
\def\r@fcite#1,{\ifuncit@d{#1}
    \newr@f{#1}%
    \expandafter\gdef\csname r@ftext\number\r@fcount\endcsname%
                     {\message{Reference #1 to be supplied.}%
                      \writer@f#1>>#1 to be supplied.\par}%
 \fi%
 \csname r@fnum#1\endcsname}
\def\ifuncit@d#1{\expandafter\ifx\csname r@fnum#1\endcsname\relax}%
\def\newr@f#1{\global\advance\r@fcount by1%
    \expandafter\xdef\csname r@fnum#1\endcsname{\number\r@fcount}}

\let\r@fis=\refis			
\def\refis#1#2#3\par{\ifuncit@d{#1}
   \newr@f{#1}%
  \w@rnwrite{Reference #1=\number\r@fcount\space is not cited up to now.}\fi%
  \expandafter\gdef\csname r@ftext\csname r@fnum#1\endcsname\endcsname%
  {\writer@f#1>>#2#3\par}}

\def\ignoreuncited{
   \def\refis##1##2##3\par{\ifuncit@d{##1}%
    \else\expandafter\gdef\csname r@ftext\csname r@fnum##1\endcsname\endcsname%
     {\writer@f##1>>##2##3\par}\fi}}

\def\r@ferr{\endreferences\errmessage{I was expecting to see
\noexpand\endreferences before now;  I have inserted it here.}}
\let\r@ferences=\references
\def\references{\r@ferences\def\endmode{\r@ferr\par\endgroup}}

\let\endr@ferences=\endreferences
\def\endreferences{\r@fcurr=0
  {\loop\ifnum\r@fcurr<\r@fcount
   \advance\r@fcurr by 1\relax\expandafter\r@fis\expandafter{\number\r@fcurr}%
    \csname r@ftext\number\r@fcurr\endcsname%
  \repeat}\gdef\r@ferr{}\global\r@fcount=0\endr@ferences}

\let\r@fend=\endpaper\gdef\endpaper{\ifr@ffile
\immediate\write16{Cross References written on []\jobname.REF.}\fi\r@fend}

\catcode`@=12

\citeall\refto		
\citeall\ref		%
\citeall\Ref		%

\referencefile

\def\frac#1/#2{#1 / #2}
\def\neuphys{Department of Physics\\Northeastern University\\Boston MA 02115}
\def\mitctp{Center for Theoretical Physics\\
Massachusetts Institute of Technology\\Cambridge MA 02139}
\def\oneandfourfifthsspace{\baselineskip=\normalbaselineskip
  \multiply\baselineskip by 9 \divide\baselineskip by 5}

\font\titlefont=cmr10 scaled\magstep3
\def\bigtitle                      
  {\null\vskip 3pt plus 0.2fill
   \beginlinemode \doublespace \raggedcenter \titlefont}

\def\Dplus{{\bf\bar{\mit D}}}
\def\Dminus{{D}}
\def\ubar{{\bf\bar{\mit u}}}
\def\dbar{{\bf\bar {\mit d}}}
\def\ebar{{\bf\bar {\mit e}}}
\def\nubar{{\bf\bar {\mit \nu}}}

\oneandfourfifthsspace
\preprintno{MIT-CTP-2345}
\preprintno{NUB-3097-94TH}
\preprintno{hep-ph/9408230}
\bigtitle{Discrete symmetries and isosinglet quarks
in low-energy supersymmetry}
\medskip
\author Diego J. Casta\~no
\affil\mitctp
\smallskip
\centerline{\rm and}
\author
Stephen P. Martin$^\dagger$
\affil\neuphys
\body

\footnote{}{$^\dagger$ Address after Sept.~1, 1994: Randall Physics
Laboratory, University of Michigan, Ann Arbor MI 48109}

\abstract
Many extensions of the minimal supersymmetric standard model contain
superfields for quarks which are singlets under weak isospin with electric
charge $-1/3$. We explore the possibility that such isosinglet quarks have
low or intermediate scale masses, but do not mediate rapid proton decay
because of a discrete symmetry. By imposing the discrete gauge anomaly
cancellation conditions, we show that the simplest way to achieve this is to
extend the $Z_3$ ``baryon parity" of Ib\'a\~nez and Ross to the isosinglet
quark superfields. This can be done in three distinct ways. This strategy is
not consistent with grand unification with a simple gauge group, but may find
a natural place in superstring-inspired models, for example. An interesting
feature of this scenario is that proton decay is absolutely forbidden.

\endtitlepage

Extensions of the standard model with supersymmetry unbroken down to energies
comparable with the electroweak-breaking scale can solve the naturalness
problem associated with the Higgs scalar boson. It is remarkable that in
the minimal supersymmetric standard model [\cite{reviews}] (MSSM), the three
gauge couplings appear to unify [\cite{unification}] at a scale
$\sim 10^{16}$ GeV, hinting at a grand unified theory (GUT) or some other
organizing principle such as superstring theory. There is a potential
phenomenological embarrassment in such theories, however; they
contain chiral superfields for quarks which are singlets of weak
isospin and carry electric charge $-1/3$. In GUT models,
these isosinglet quarks
necessarily appear in the same multiplets as the Higgs doublets
of the MSSM, so that generically one might expect them to have masses
comparable to the electroweak scale. In superstring-inspired models, the
chiral superfields come from remnants of the ${\bf 27}$ and
$\overline{\bf 27}$ representations of $E_6$. The masses of the isosinglet
quark superfields are extremely model-dependent, and are typically determined
by perturbations from flat directions in the superpotential. Therefore,
again in superstring models, the isosinglet quarks can very
often have low or intermediate scale masses.

The most general superpotential for the MSSM plus the isosinglet quarks
is given schematically by:
$$
\eqalign{
W & = W_0 + W_1 + W_2
\cr
W_0 & = Q H_u \ubar + Q H_d \dbar + L H_d \ebar
+ \mu H_u H_d + \mu_D  \Dminus \Dplus
\cr
W_1 & = QL\Dplus + \ubar \ebar \Dminus
\cr
W_2 & = QQ\Dminus +  \ubar \dbar \Dplus
\> .\cr
}
$$
Here $Q,\ubar,\dbar,L,\ebar$ are the quark and lepton chiral superfields
of the MSSM; $H_u,H_d$ are the Higgs doublet chiral superfields
of the MSSM; and $\Dminus,\Dplus$ are the chiral
superfields for the isosinglet quarks. Under the gauge group
$SU(3)_c \times SU(2)_L \times U(1)_Y$, they transform as
$$
{\eqalign{
Q &\sim ({\bf 3},{\bf 2},{1/6})
\cr
L &\sim ({\bf 1},{\bf 2},-{1/2})
\cr
H_d &\sim ({\bf 1},{\bf 2},-{1/2})
\cr }}
\qquad\qquad
{\eqalign{
\ubar &\sim (\overline{\bf 3},{\bf 1},-{2/3})
\cr
\ebar &\sim ({\bf 1},{\bf 1},1)
\cr
\Dminus &\sim ({\bf 3},{\bf 1},-{1/3})
\cr }}
\qquad\qquad
{\eqalign{
\dbar &\sim (\overline{\bf 3},{\bf 1},{1/3})
\cr
H_u &\sim ({\bf 1},{\bf 2},{1/2})
\cr
\Dplus &\sim (\overline{\bf 3},{\bf 1},{1/3})
\> .
\cr }}
$$
For now, we assume the conservation of the usual $Z_2$ matter parity
which is given for each chiral superfield in the MSSM by $(-1)^{3(B-L)}$, where
$B$ and $L$ are the usual baryon number and total lepton number.
This matter parity is trivially related to R-parity
by a minus sign for fermions.
Thus $Q,\ubar,\dbar,L,\ebar$ all have matter parity $-1$, and
$H_u,H_d,\Dminus,\Dplus$
each have matter parity $+1$. Later we will consider the implications of
relaxing this assumption. We assume that there are 3 chiral
families of quarks and leptons, and the isosinglet quark superfields
$\Dminus$ and $\Dplus$ may or may not also be replicated, but we suppress
all flavor and gauge indices.

The most pressing phenomenological problem posed by the existence of
isosinglet quarks is the possibility of rapid proton decay. For example,
if both of the terms $QL\Dplus$ and $QQ\Dminus$ existed in the superpotential
with couplings of order unity,
and $\Dminus,\Dplus$ had a mass $\mu_D$ in the TeV range, then the proton
would decay in minutes
due to one-loop diagrams with the virtual exchange of an isosinglet quark and
a wino. The dominant decay mode would be $p\rightarrow K^+{\bar\nu}$,
for which Kamiokande has established the experimental limit [\cite{Kamiokande}]
$\tau(p\rightarrow K^+{\bar\nu}) > 10^{32}$ yrs.
More generally, the presence of
either term in $W_1$ together with either term from $W_2$ will prevent us from
consistently assigning $B$ or $L$ to the chiral superfields in the theory,
generically resulting in catastrophic proton decay.
If the isosinglet quarks exist at all, then there appear to be two ways out of
this disaster; either $\Dminus$ and $\Dplus$  must be very heavy so that
their effects on low energy physics very nearly decouple, or some
additional symmetry must be invoked to explain why either $W_1$ or $W_2$
or both are missing.

Both of these potential solutions are problematic in a supersymmetric GUT.
It is possible to arrange for $\Dminus$ and $\Dplus$
to obtain a very large mass;
however, this requires some cleverness because at least one copy of
$\Dminus$ and $\Dplus$ lives in
the same multiplet of the GUT gauge group as $H_u$ and $H_d$.
Various proposals have been put forward to effect a separation in mass
scales between $\Dminus,\Dplus$ and $H_u,H_d$, including the
``sliding singlet" mechanism [\cite{slidingsinglet}], the ``missing partner"
mechanism [\cite{missingpartner}], the related ``missing VEV" mechanism
[\cite{DW}], and Higgses as Nambu-Goldstone bosons [\cite{ngbosons}].
These attempts generally require an intricate system of global symmetries
to provide for realistic quark and lepton masses.
It is also possible in supersymmetric
GUTs to accept light isosinglet quarks but to rely on delicate cancellations
among couplings to prevent proton decay [\cite{BR}].

In this paper we consider instead the possibility that isosinglet
quarks $\Dminus,\Dplus$ are light, but a discrete symmetry prohibits the
terms from either $W_1$ or $W_2$ or both. This strategy
is not consistent with a GUT based on a simple gauge group,
but could be useful in superstring-inspired models, for example.
The possibilities may then be divided into three cases, as follows:

\noindent {\bf Case A}:
$QL\Dplus$ and $\ubar\ebar\Dminus$ are
allowed; $QQ\Dminus$ and $\ubar\dbar\Dplus$ are forbidden. Then we can
assign baryon number and lepton number $B=1/3$, $L=1$ to $\Dminus$ and
$B=-1/3$, $L=-1$ to $\Dplus$. Thus $\Dminus$ and $\Dplus$ are ``leptoquarks".

\noindent {\bf Case B}:
$QQ\Dminus$ and $\ubar\dbar\Dplus$  are
allowed; $QL\Dplus$ and $\ubar\ebar\Dminus$ are forbidden. Then we can
assign baryon number and lepton number   $B=-2/3$, $L=0$ to $\Dminus$  and
$B=2/3$, $L=0$ to $\Dplus$. Thus $\Dminus$ and $\Dplus$ are ``diquarks".

\noindent {\bf Case C}:
$QQ\Dminus$, $\ubar\dbar\Dplus$, $QL\Dplus$, and $\ubar\ebar\Dminus$ are
all forbidden. Here it is not yet clear how to assign $B$ and $L$
to the isosinglet quark superfields, since they have no renormalizable
superpotential interactions other than a mass term.
\phantom{\cite{DT,Kizukuri,BDH,AEKNTZ,Ma,MN,HR}}
\phantom{\cite{KW,IR1,IR2,BD,moreibanez}}

There has already been much interest [\cite{DT}-\cite{HR}]
in the phenomenological implications of each of these three cases, particularly
in the context of superstring models based on remnants of $E_6$. In this paper
we will examine possibilities for discrete symmetries which can enforce the
missing couplings in each of the three cases. Specifically, we consider
a $Z_N$ symmetry under which each chiral superfield transforms as
$$
\Phi \rightarrow {\rm exp}{(2 \pi i \alpha_\Phi /N)} \Phi
\eqno(zn)
$$
where the $\alpha_\Phi$ are the additive $Z_N$ charges. An operator is allowed
if and only if the sum of its $Z_N$ charges is 0 [mod $N$]. For simplicity,
we will assume that the $Z_N$ charges are not family-dependent; this seems to
be required for the quark superfields anyway in order to allow for the observed
Cabibbo-Kobayashi-Maskawa mixing. Since even a small violation of the discrete
symmetry could result
in catastrophic proton decay, it is strongly suggested that the $Z_N$ is a
``gauged" discrete symmetry. One way [\cite{KW}]
(but perhaps not the only way) to obtain
a gauged discrete symmetry is to break a gauged $U(1)$ symmetry with an order
parameter whose charge is $Nq$, where the smallest non-zero
$U(1)$ charge assignment
in the theory is $q$. Unlike a global symmetry, a gauged discrete symmetry
is automatically protected against violation by Planck-scale and
other non-perturbative effects. As shown in [\cite{IR1}-\cite{moreibanez}],
such gauged discrete symmetries
are subject to stringent requirements based on anomaly cancellation.

Requiring that the $Z_N$ symmetry allow
the usual Yukawa couplings and masses in $W_0$,
one immediately obtains some relations between the $Z_N$-charges. Thus
$$
\alpha_{H_d} = - \alpha_{H_u}
\qquad {\rm and} \qquad \alpha_\Dplus = - \alpha_\Dminus
\eqno(allowmass)
$$
are required in order to allow for Higgs and isosinglet quark masses
respectively and
$$
\alpha_\ubar = -\alpha_Q - \alpha_{H_u} ,
\qquad
\alpha_\dbar = -\alpha_Q + \alpha_{H_u} ,
\qquad
\alpha_\ebar = -\alpha_L + \alpha_{H_u} .
\eqno(allowyuks)
$$
in order to allow for Yukawa couplings.
Each of these equations is understood to hold modulo $N$.
We get further constraints [mod $N$] in each of the three cases:

\noindent { Case A}:~~~~$\alpha_\Dminus = \alpha_Q + \alpha_L$
and $3\alpha_Q + \alpha_L \not= 0$.

\noindent { Case B}:~~~~$\alpha_\Dminus = -2 \alpha_Q $ and again
$3\alpha_Q + \alpha_L \not= 0$.

\noindent { Case C}:~~~~$\alpha_\Dminus \not= \alpha_Q +\alpha_L$ and
$\alpha_\Dminus \not= - 2 \alpha_Q $.

\noindent Now, cases A and B are in some sense more interesting,
because in case C the isosinglet quarks are relatively sterile,
having only gauge interactions with the chiral superfields of the MSSM.
Therefore, we consider cases A and B first.

To further constrain the $Z_N$ symmetry, we now consider the discrete
anomaly cancellation conditions of Ib\'a\~nez and Ross [\cite{IR1}].
First we consider
the mixed $Z_N\times SU(n) \times SU(n)$ anomaly cancellation condition,
which is given in general by
$$
2 \sum_i \alpha_i T_i  = 0 \qquad\qquad [{\rm mod}\>N]
$$
where $T_i$ is the $SU(n)$ Dynkin index for each representation,
normalized so that the fundamental representation
has $T=1/2$. By plugging in the
constraints already considered, it is easy to show that the
$Z_N \times SU(3)_c \times SU(3)_c$ anomaly always cancels. However,
the $Z_N \times SU(2)_L \times SU(2)_L$ anomaly cancellation condition
is given by
$$
n_f (3 \alpha_Q + \alpha_L) = 0 \qquad\qquad [ {\rm mod}\> N ]
\eqno(n22)
$$
where $n_f=3$ is the number of chiral MSSM families.
This is quite non-trivial, since we found that in both cases A and B,
$$
3 \alpha_Q + \alpha_L \not= 0 \qquad\qquad [ {\rm mod}\> N ]
\eqno(constraint)
$$
in order to allow only the appropriate interactions for $\Dminus$ and $\Dplus$.
The only way to satisfy both of these constraints simultaneously is
to take $N$ to be a multiple of 3, since that is the number of families.
While larger values of $N$ certainly might give interesting models,
we will take $N=3$ in the following for simplicity; it is worth noting
that larger discrete symmetries are harder to obtain in superstring models.

The mixed gravitational ($Z_N\times G \times G$) anomaly
cancellation condition is given in general by
$$
\sum_i \alpha_i = 0
\qquad\qquad
{ [{\rm mod} \> N] \>\> (N \> {\rm odd})
\atop
[{\rm mod} \> { N/2}] \>\> (N \> {\rm even})\> .  }
$$
This gives no further constraint for $N=n_f=3$.
Finally we consider the $Z_N^3$ anomaly cancellation condition, which
is given in general by
$$
\sum_i \alpha_i^3 = r N + \eta s N^3/8,
$$
where $\eta = 0$(1) for $N$ odd (even) and $r,s$ are integers, and for $N=3$,
$r$ must in fact be a multiple of 3 [\cite{IR1}].
Given the relations already found, the cubic anomaly then
cancels provided that $\alpha_{H_u} = -\alpha_L$.
The $Z_N^3$ anomaly cancellation
condition actually can {\it always} be satisfied
for any choice of charge assignments [\cite{BD}], but in our case
this would require
the existence of heavy fields with fractional $Z_3$ charges, so that the
underlying discrete symmetry would have to be at least as large as $Z_9$
[\cite{moreibanez}].

Now, one can always take $\alpha_Q=0$, by redefining the $Z_N$ symmetry
according to
$$
\alpha_\Phi \rightarrow \alpha_\Phi + 6 n Y_\Phi
\eqno(shift)
$$
where $n$ is an appropriate integer and $Y_\Phi$ is the weak hypercharge
of $\Phi$. From eq.~\(constraint), $\alpha_L$ cannot vanish, so
we can also choose $\alpha_L=1$ without loss of generality.
In our case, this completely fixes all
of the $Z_3$ charges. For the chiral superfields of the MSSM, we have
$$
\alpha_Q = 0; \qquad\! \alpha_L = 1; \qquad\! \alpha_\ubar = 1;
\qquad\! \alpha_\dbar = -1;
\qquad\! \alpha_\ebar = 1;\qquad\! \alpha_{H_u} = -1;\qquad\! \alpha_{H_d} = 1
\> .
$$
For the isosinglet quark superfields
there are now three possibilities, which in fact just
correspond to the three cases described above:

\noindent Case A:~~~~$\alpha_\Dminus = 1; \qquad \alpha_\Dplus = -1$,

\noindent Case B:~~~~$\alpha_\Dminus = 0; \qquad \alpha_\Dplus = 0$,

\noindent Case C:~~~~$\alpha_\Dminus = -1; \qquad \alpha_\Dplus = 1$ .

\noindent Thus we have uniquely fixed the charges of the chiral superfields
under the discrete symmetries $Z_3^A$ and $Z_3^B$  for cases
A and B respectively, and found a $Z_3^C$ for case C whose MSSM charge
assignments are consistent with cases A and B. [Of course, in case C only,
$N$ need not be a multiple of 3 since \(constraint) need not hold. If one were
only interested in the sterile $\Dminus, \Dplus$ of
case C, one could have a consistent solution for any $N$
by, e.g., assigning $Z_N$ charges $\alpha_\Dminus = -\alpha_\Dplus = 1$
and $\alpha_i =0$ for all MSSM superfields.]

Note that each of $Z_3^A$, $Z_3^B$, and $Z_3^C$, when restricted to
the fields of the MSSM, give  nothing
other than the discrete ``baryon parity" proposed by Ib\'a\~nez and Ross
[\cite{IR2}] as
an alternative to the usual matter parity, and which is given in their
classification scheme by $R_3 L_3$. They showed that this is the only
family-independent $Z_3$ discrete symmetry with the minimal particle content
which satisfies the discrete anomaly cancellation conditions.
Its reappearance here is not at all surprising,
since the existence of the isosinglet quarks cannot affect the anomaly
cancellation
conditions, due to the constraint that $\Dminus$ and $\Dplus$ must have
opposite charges to allow a mass term in the superpotential.

It is natural to consider the extension of this model to include
gauge-singlet neutrino chiral superfields $\nubar$ in order to provide for a
realistic and non-trivial
neutrino mass spectrum via the seesaw mechanism [\cite{seesaw}].
Thus we may consider adding to the usual superpotential terms of the form
$$
W_{\rm seesaw} = L H_u \nubar + M_\nubar \nubar\nubar \> .
$$
Both of these terms are allowed if we assign $\alpha_\nubar = 0$. Then the
anomaly cancellation conditions are not affected, and the term
$\nubar\dbar\Dminus$, which is naturally associated with $W_1$, is allowed
in case A and forbidden in cases B and C.

In the scenario described here, the usual $Z_2$ matter parity which is
equivalent to R-parity is actually
optional. If one does not impose matter parity,
the superpotential might {\it a priori} include $B$-violating terms
$$
W_3 = \ubar\dbar\dbar
$$
and $L$-violating terms
$$
W_4 = QL\dbar + LL\ebar + L H_u   \> .
$$
However, $Z_3^{A,B,C}$ necessarily forbid $\ubar\dbar\dbar$,
so that there can be no $B$ violation among renormalizable interactions.
In fact, the $Z_3$ charges as given above are equal, for each chiral
superfield, to $B- 2Y$ where $Y$ is the weak hypercharge. Since $Y$ is
conserved, $Z_3^{A,B,C}$ invariance implies the selection rule
$$
\Delta B = 0 \qquad\qquad [{\rm mod}\> 3]
\eqno(selectionrule)
$$
for operators of arbitrarily high dimension!
Therefore, proton decay is {\it absolutely} forbidden
by $Z_3^{A,B,C}$. Likewise, neutron--anti-neutron
oscillations are absolutely forbidden by this same selection rule, since
they would violate $B$ by 2 units.
Soft supersymmetry-breaking interactions do not
affect this conclusion, because they respect the same symmetries as the
terms in the superpotential.
Although $B$-violating effects are highly constrained by the selection rule
\(selectionrule), baryogenesis can still occur if a lepton number asymmetry
is partially transformed into a baryon number asymmetry by high temperature
electroweak effects.

Note that $\langle H_u \rangle $ and  $\langle H_d \rangle $
transform non-trivially under $Z_3^{A,B,C}$, producing the spontaneous
symmetry breaking
$$
SU(2)_L \times U(1)_Y \times Z_3^{A,B,C}
\rightarrow U(1)_{EM} \times Z_3^{\prime A,B,C}
$$
at the electroweak scale. The $Z_3^{\prime A,B,C}$ charges are
given for each particle by $B+Q_{EM}$, where $Q_{EM}$
is the electric charge. Thus $B$ violation is still governed by the
selection rule \(selectionrule), as should be clear from the fact that
the $H_u$ and $H_d$ carry no baryon number.

On the other hand, all of the lepton-number
violating interactions in $W_4$ are allowed by the discrete symmetries
$Z_3^{A,B,C}$. If R-parity is imposed in addition to the $Z_3^{A,B,C}$
symmetry, then $L$ will be conserved for all renormalizable interactions
except $W_{\rm seesaw}$.
Also, imposing R-parity will forbid mixing between $\dbar$ and
$\Dplus$ in case A,
which may be important in avoiding flavor-changing neutral currents.
If R-parity is not imposed in addition to $Z_3^{A,B,C}$,
there will be dramatic effects on accelerator searches for supersymmetry,
since the LSP may decay into leptonic states inside the detector, and
sleptons will have direct two-body decays into lepton+neutrino.

The fermionic components of $\Dminus,\Dplus$ have a Dirac mass, $\mu_D$.
Their scalar partners
will also receive contributions to their masses from soft
supersymmetry-breaking terms. The isosinglet quarks of cases A and B
can readily decay to MSSM states.
In case A, a possible decay signature for the scalar leptoquark is a jet and
either a lepton or missing energy.  The heavier of the scalar and fermionic
leptoquark can also decay to the lighter and a neutralino
[\cite{BDH}].  The fermionic leptoquark can also decay semileptonically
to two- or three- body states, depending on the kinematics.  In case B,
the direct scalar diquark two-jet decay will be difficult to see because of
comparatively large QCD backgrounds [\cite{AEKNTZ},\cite{HR}].  The
fermionic diquark may decay through interesting two- or three- body
channels, depending on the sparticle  mass spectrum.

In case C, there is a phenomenological danger from the fact that the isosinglet
quarks cannot decay into MSSM particles and might therefore be too stable.
If R-parity is not conserved, there is an additional renormalizable interaction
available to the isosinglet quarks in Case C, namely
$$
[\ubar\Dplus\Dplus]_F         \eqno(uDD)
$$
which seems to imply the curious assignment $B=-1/6$ and $L=0$ for $\Dminus$,
although one could also assign $B=4/3$, $L=0$ to $D$, so that
\(uDD) violates $B$ by 3 units in agreement with the selection rule
\(selectionrule).
This operator is forbidden by $Z_3^A$ and $Z_3^B$, and it
vanishes unless there is more than one copy of $\Dplus$.
In any case, it does not help the lightest of the isosinglet (s)quarks to
decay. If it is
forbidden, then the renormalizable interactions do not fix $B$ or $L$ for the
isosinglet quarks. There is a possible non-renormalizable operator which can
allow the isosinglet quarks to decay, namely the dimension-6 operator
$$
[\ubar\dbar\dbar\dbar\Dminus]_F    \> .     \eqno(dim6)
$$
This is in fact the only operator of dimension 6 or less
which is allowed by $Z_3^C$ and which
can provide for isosinglet (s)quarks decaying into standard model states.
If this provides the dominant decay mode for the isosinglet quarks, then
we are led to the assignments $B=4/3$, $L=0$ for $\Dminus$ in case
C. This dimension-6 operator is suppressed by two powers of some putative
high mass scale $M_X$, and leads to a many-body decay with lifetime
$\sim M_X^4/\mu_D^5$.  So the case C isosinglet quark may be long-lived and
cosmologically dangerous if $M_X \gg \mu_D$, although this depends crucially
on the nature of the new physics at the scale $M_X$, about which we will
not speculate here. Also, it is conceivable that
particles not considered here which are lighter than
$\Dminus,\Dplus$ might provide additional decay channels in case C.

Of course,
the presence of light or intermediate scale $\Dminus,\Dplus$ will affect
the running of the gauge couplings, in general ruining the unification of
gauge couplings and washing out the successful ``prediction"
of $\sin^2\theta_W$. However, the addition of other thresholds
(due to vectorlike particles at intermediate scales) might easily
restore the correct
value of $\sin^2\theta_W$, perhaps with a true gauge coupling unification
scale which is higher than the apparent unification
scale at $\sim 10^{16}$ GeV.
Realistic superstring models often have several such intermediate thresholds.
As we have already mentioned, the strategy used
in this paper is not consistent
with a GUT. This is true even if one takes the GUT multiplet partners of
the MSSM Higgs to be very heavy and does not identify them with $\Dminus$ and
$\Dplus$, because the $Z_3$ charges of the MSSM quark and lepton superfields
are not consistent with assigning them to GUT multiplets.

Cancellation of discrete anomalies might also occur in
superstring models through a discrete version of the
Green-Schwarz (GS) mechanism [\cite{gsmechanism}].  This may occur if
the axion which is the partner of the dilaton transforms non-trivially
under the discrete gauge group. However, the discrete GS mechanism does
not allow any new solutions to the constraints given above in the
usual case that the Kac-Moody levels $k_2$ and $k_3$ of
the gauge groups $SU(2)_L$ and $SU(3)_c$ are equal. To see this,
suppose we have a discrete $Z_N$ symmetry which satisfies \(allowmass)
and \(allowyuks). With a discrete GS mechanism, the
$Z_N \times SU(2)_L \times SU(2)_L$ and
$Z_N \times SU(3)_c \times SU(3)_c$ mixed anomaly cancellation conditions
become
$$
\eqalignno{
n_f (3 \alpha_Q + \alpha_L) & = 2 \delta_{\rm GS} k_2 \qquad [{\rm mod}\> N]
&(gs2)
\cr
0 & = 2 \delta_{\rm GS} k_3 \qquad [{\rm mod}\> N]
&(gs3)
\cr
}
$$
respectively, where $\delta_{\rm GS}$ is a constant.
But now from \(gs3) we immediately find that if $k_3=k_2$,
then \(gs2) just reduces to the previous constraint \(n22).

There have been previous attempts to implement discrete symmetries
to remove or suppress proton decay from light color triplets in
superstring-inspired models. However, these preceded the appearance of
[\cite{IR1}] and so do not
take into account discrete anomaly cancellation. Reference
[\cite{DT}] considers a $Z_2$ symmetry which is somewhat
similar to our case A, but which has a
$Z_2 \times SU(2)_L \times SU(2)_L$ anomaly, and also a
$Z_2$ symmetry which is discrete anomaly-free and corresponds to our case
C, but which requires massless neutrinos. They also consider a
family-dependent $Z_3$ symmetry intended to suppress $\mu \rightarrow e
\gamma$, but which has a $Z_3 \times SU(2)_L \times SU(2)_L$ anomaly.
In references [\cite{Kizukuri}] and [\cite{Ma}], other $Z_2$
discrete symmetries corresponding to our cases A and B are proposed,
but they have $Z_2 \times SU(2)_L \times SU(2)_L$ anomalies,
as we have shown on general grounds.

In this paper, we have found that light weak-isosinglet quark superfields
can be added to the particles of the MSSM without causing rapid proton
decay, if the theory is invariant under a discrete symmetry. By using
the  discrete gauge anomaly cancellation conditions, we showed that
by far the most economical way to do this is to extend the ``baryon
parity" of [\cite{IR2}] to the isosinglet quark superfields. In fact,
each of the three possible ways to do this can lead to acceptable
phenomenology. Note that
in general, one can have isosinglet quarks for each of cases A, B, and C
peacefully coexisting, since the restrictions of $Z_3^{A,B,C}$ to the
MSSM chiral superfields are consistent.
The three classes of isosinglet quarks will not
mix with each other, since they have different $Z_3$ charges. Even if there
are no isosinglet quarks near the TeV scale, the $Z_3$ symmetry may be useful
for preventing proton decay in models with $\Dminus,\Dplus$
at intermediate scales.
An intriguing theoretical aspect of the $Z_3$ symmetry is that its
consistency is tied to the presence of 3 chiral families in the MSSM
through the discrete anomaly cancellation conditions.
Finally, we note that the scenario described here is falsifiable,
since the selection rule \(selectionrule) forbids all proton decay.
Future proton decay searches at Super Kamiokande
and ICARUS will therefore be crucial tests.

\noindent Acknowledgments:
We are grateful to Pierre Ramond for helpful comments.
The work of D.J.C. was supported in part by the U.S. Department of Energy
under cooperative agreement DE-FC02-94ER40818 and in part by the
Texas National Research Laboratory Commission under grant RGFY93278C.
The work of S.P.M. was supported in part by the
National Science Foundation grants PHY-90-01439 and PHY-93-06906 and
U.~S.~Department of Energy grant DE-FG02-85ER40233.

\references

\refis{BR} M.~J.~Bowick and P.~Ramond, \pl 131B, 367, 1983.

\refis{unification} P.~Langacker, in Proceedings of the
PASCOS90 Symposium, Eds.~P.~Nath
and S.~Reucroft, (World Scientific, Singapore 1990)
J.~Ellis, S.~Kelley, and D.~Nanopoulos, \pl 260B, 131, 1991;
U.~Amaldi, W.~de Boer, and H.~Furstenau, \pl 260B, 447, 1991;
P.~Langacker and M.~Luo, \pr D44, 817, 1991.

\refis{KW} L.~Krauss and  F.~Wilczek, \prl 62, 1221, 1989.

\refis{reviews}
For reviews, see H.~P.~Nilles,  \prpts 110, 1, 1984 and
H.~E.~Haber and G.~L.~Kane, \prpts 117, 75, 1985.

\refis{BD} T.~Banks and M.~Dine, \pr D45, 1424, 1992.

\refis{IR1} L.~E.~Ib\'a\~nez and  G.~G.~Ross, \pl B260, 291, 1991.

\refis{IR2} L.~E.~Ib\'a\~nez and  G.~G.~Ross, \np B368, 3, 1992.

\refis{seesaw} M.~Gell-Mann, P.~Ramond, and R.~Slansky, in Sanibel Talk,
CALT-68-709, Feb 1979, and in {\it Supergravity},
(North Holland, Amsterdam, 1979; T.~Yanagida, in {\it Proc. of the
Workshop on
Unified Theories and Baryon Number in the Universe}, Tsukuba, Japan, 1979,
edited by A. Sawada and A. Sugamoto (KEK Report No. 79-18, Tsukuba, 1979).

\refis{slidingsinglet} E.~Witten, \pl B105, 267, 1981;
L.~E.~Ib\'a\~nez and G.~G.~Ross, \pl B110, 215, 1982.

\refis{missingpartner}
B.~Grinstein, \np B206, 387, 1982;
A.~Masiero, D.~V.~Nanopoulos,
K.~Tamvakis, and T.~Yanagida, \pl B115, 380, 1982.

\refis{Kamiokande} Kamiokande collaboration, K.~S.~Hirata et.~al.,
\pl B220, 308, 1989.

\refis{DW}
S.~Dimopoulos and F.~Wilczek, preprint NSF-ITP-82-07 (unpublished);
K.~S.~Babu and S.~M.~Barr, \pr D48, 5354, 1993.

\refis{ngbosons} K.~Inoue, A.~Kakuto, and T.~Takano, \journal
Prog. Theor. Phys., 75, 664, 1986;
A.~Anselm and A.~Johansen, \pl B200, 331, 1988;
R.~Barbieri, G.~Dvali, and A.~Strumia, \np B391, 487, 1993.

\refis{DT} M.~Drees and X.~Tata, \prl 59, 528, 1987.

\refis{Kizukuri} Y.~Kizukuri, \pl B185, 183, 1987.

\refis{BDH} V.~Barger, N.~G.~Deshpande, and K.~Hagiwara,
\pr D36, 3541, 1987.

\refis{AEKNTZ} V.~D.~Angelopoulos, J.~Ellis, H.~Kowalski,
D.~V.~Nanopoulos, N.~D.~Tracas, and F.~Zwirner, \np B292, 59, 1987.

\refis{Ma} E.~Ma, \prl 60, 1363, 1988.

\refis{MN} E.~Ma and D.~Ng, \pr D39, 1986, 1989.

\refis{HR} J.~L.~Hewett and T.~G.~Rizzo, \prpts 183, 193, 1989.

\refis{gsmechanism} M.~B.~Green and J.~H.~Schwarz, \pl B149, 117, 1984.

\refis{moreibanez} L.~E.~Ib\'a\~nez, \np B398, 301, 1993.

\endreferences\endit\end